\documentclass[twocolumn]{aastex62}

\newcommand{\wcen}{$\omega\ \rm{Cen}$}

\shorttitle{Discovery of Millisecond Pulsars in the Globular Cluster Omega Centauri}
\shortauthors{Dai et al.}

\begin{document}

\title{Discovery of Millisecond Pulsars in the Globular Cluster Omega Centauri}

\correspondingauthor{Shi Dai}
\email{shi.dai@csiro.au}

\author[0000-0002-9618-2499]{Shi Dai}
\affiliation{CSIRO Astronomy and Space Science, Australia Telescope National Facility, Epping, NSW 1710, Australia}

\author[0000-0002-7122-4963]{Simon Johnston}
\affiliation{CSIRO Astronomy and Space Science, Australia Telescope National Facility, Epping, NSW 1710, Australia}

\author[0000-0002-0893-4073]{Matthew Kerr}
\affiliation{Space Science Division, Naval Research Laboratory, Washington, DC 20375-5352, USA}

\author[0000-0002-1873-3718]{Fernando Camilo}
\affiliation{South African Radio Astronomy Observatory, Observatory 7925, South Africa}

\author[0000-0002-2037-4216]{Andrew Cameron}
\affiliation{CSIRO Astronomy and Space Science, Australia Telescope National Facility, Epping, NSW 1710, Australia}

\author{Lawrence Toomey}
\affiliation{CSIRO Astronomy and Space Science, Australia Telescope National Facility, Epping, NSW 1710, Australia}

\author{Hiroki Kumamoto}
\affiliation{Kumamoto University, Graduate School of Science and Technology, Japan}
\affiliation{CSIRO Astronomy and Space Science, Australia Telescope National Facility, Epping, NSW 1710, Australia}


\begin{abstract}
The globular cluster Omega Centauri is the most massive and luminous cluster in the Galaxy. The $\gamma$-ray source FL8Y~J1326.7--4729 is coincident with the core of the cluster, leading to speculation that hitherto unknown radio pulsars or annihilating dark matter may be present in the cluster core. Here we report on the discovery of five millisecond pulsars in Omega Centauri following observations with the Parkes radio telescope. Four of these pulsars are isolated with spin periods of 4.1, 4.2, 4.6 and 6.8\,ms. The fifth has a spin period of 4.8\,ms and is in an eclipsing binary system with an orbital period of 2.1 hours. Deep radio continuum images of the cluster centre with the Australian Telescope Compact Array reveal a small population of compact radio sources making it likely that other pulsars await discovery. We consider it highly likely that the millisecond pulsars are the source of the $\gamma$-ray emission. The long-term timing of these pulsars opens up opportunities to explore the dynamics and interstellar medium of the cluster.

\end{abstract}


\section{Introduction} 
\label{sec:intro}
Of the more than 200 globular clusters (GCs) known in the Milky Way, Omega Centauri (\wcen\ or NGC~5139) stands out. It is not only the most massive and luminous GC in the Galaxy, but also shows characteristic features, such as its very broad metallicity distribution~\citep[e.g.,][]{fr75,msb+19} 
and its incredible multiplicity in stellar populations~\citep[e.g.,][]{pfb+00,bma+17}.
\wcen\ has one of the largest cores with an angular radius of 155\,arcsec~\citep{har10}, and a projected number density of $\sim10^6$ stars in the central region with high velocity dispersion~\citep{av10}.
The properties of \wcen\ have led to suggestions that it was once a dwarf galaxy captured by the Milky Way with its outer stellar envelope almost entirely removed by tidal stripping~\citep[e.g.,][]{bf03}.

The $\gamma$-ray source FL8Y~J1326.7--4729 \citep{aaa+10} is coincident with the core of \wcen. Discovered shortly after the launch of the Fermi satellite, the hard spectrum and exponential cut-off are very typical of emission from millisecond pulsars (MSPs). In addition, \cite{aaa+10} concluded that some $19\pm9$ pulsars could reside in the cluster. More recent analysis of nine years of Fermi data predicted a similar number of MSPs in \wcen~\citep{dcn19}. A large population of X-ray sources in the cluster core led \cite{hch+18} to conclude that MSPs were likely responsible for some of them.

Two further questions, triggered by \wcen's unique formation history and properties, have drawn much attention. First, are there any intermediate-mass black holes (IMBHs) in the centre of \wcen~\citep[e.g.,][]{ngb08,bau17}? Secondly, is there any evidence of dark matter annihilation~\citep[e.g.,][]{bml+19,kmm+19}? 
Theoretical studies and simulations of the growth rate of stars via stellar collisions in dense star clusters predict that GCs could contain IMBHs~\citep{pm02}. As the most massive GC with a large core, \wcen\ is one of the main targets to search for candidate IMBHs. While recent dynamical and kinematical studies of the central few arcmin of \wcen\ have ruled out the existence of massive IMHBs~\citep{av10,va10,bhs+19}, it is still uncertain if \wcen\ hosts an IMBH with $M_{\rm{BH}}\lesssim1.2\times10^4$\,M$_{\odot}$. Tighter constraints will be valuable for us to understand IMBH demographics in globular clusters and whether GCs have IMBHs that follow the same relationship as that established for super-massive black holes. 
On the other hand, if \wcen\ were indeed a dwarf galaxy dominated by dark matter, then its relatively close distance to Earth~\citep[$\sim5.2$\,kpc,][]{har10} makes it an excellent candidate to search for evidence of dark matter annihilation \citep{rbg+19,bml+19}. 

The discovery of radio pulsars in the core of \wcen\ could provide us with a powerful tool to solve these mysteries. As has been demonstrated for GCs 47Tuc~\citep{fkl+01,frk+17}, Terzan 5~\citep{prf+17} and NGC 6624~\citep{psl+17}, long-term timing of pulsars allows us to study the dynamics of the core and to probe any candidate IMBH and the interstellar medium. Polarisation observations of pulsars allow us to determine the rotation measure (RM), and then to estimate the strength of magnetic fields together with the pulsar dispersion measure (DM). This will be important to constrain particle dark matter annihilation models~\citep{kmm+19}. Better understanding of the MSP population will also enable us to put strong constraints on $\gamma$-rays from dark matter annihilation~\citep{rbg+19} since MSPs are also strong $\gamma$-ray sources.

Extensive searches for radio pulsars in \wcen\ have been carried out in the early 2000s~\citep{evb01,pdc+05} and more recently targeted at the Fermi Large Area Telescope (LAT) GeV source~\citep{ckr+15}, but no radio pulsars were found. Considering the size and mass of \wcen\ and the richness of MSPs in other GCs, the absence of MSPs in the cluster centre of \wcen\ was puzzling. Although this could be explained by the low rate of stellar interactions in the core of \wcen, $\gamma$-ray emission detected with Fermi-LAT suggested a small population of MSPs~\citep{aaa+10}. 

We have used the Ultra-wideband Low (UWL) receiver~\citep{hmd+19} on the Parkes radio telescope to search \wcen\ and have discovered five MSPs and have further used the Australia Telescope Compact Array to make deep continuum images. In Section~\ref{sec:observations} we present the observations and results and discuss their implications in Section~\ref{sec:discussion}.

\section{Observations and Results}
\label{sec:observations}
\subsection{Australian Telescope Compact Array Observations}
\label{sec:atca}
We observed the core of \wcen\ with the Australia Telescope Compact Array on 2019 July 10 and 2019 Sep 5 for 12 hours on each occasion at an observing frequency centred at 2.1\,GHz, with a bandwidth of 2\,GHz (CX439, PI: S. Dai). The 750C and the 6C configurations were used. The data were combined and reduced in standard fashion with the package {\sc miriad} using the flux calibrator 1934$-$63 and the phase calibrator 1320$-$446. The resultant image has a root-mean-square noise of 10.5\,$\mu$Jy and a resolution of $7.0\times4.0$~arcsec. 
In addition we obtained archival data of the cluster taken at 5.5~GHz with 2~GHz of bandwidth on 2010 Jan 22 and 23 in the 6A configuration. The image has an rms of 6.5\,$\mu$Jy and a resolution of $2.6\times1.6$~arcsec.

\begin{figure*}
\centering
\includegraphics[width=12cm]{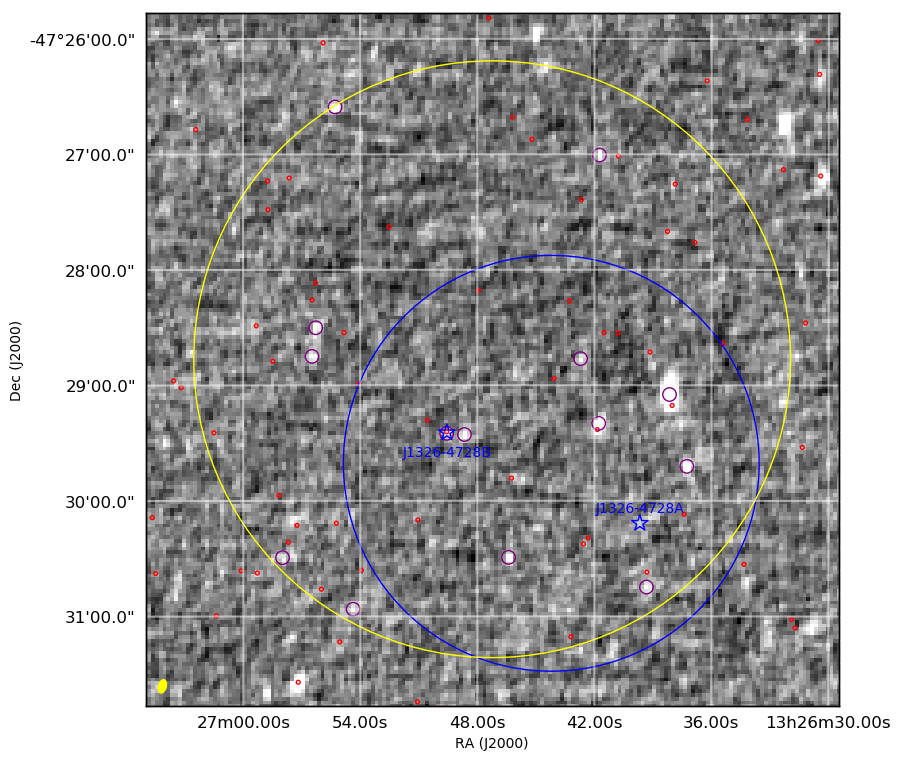}
\caption{ATCA image at 2.1GHz. The rms is 10.5\,$\mu$Jy and the resolution is 7.0 by 4.0 arcsec. The beam is shown in the bottom left corner. Purple circles: radio sources listed in Table~\ref{tab:atca} with a radius of 3.5\,arcsec. Red circles: XMM-Newton unidentified sources~\citep{hch+18} with a radius of 1\,arcsec; Blue circle: the Fermi source with radius showing its semi-major axis (107\,arcsec, 95\% confidence); Yellow circle: the core region of \wcen\ centred at RA=13:26:47.24, Dec=$-$47:28:46.5~\citep[155\,arcsec,][]{har10}. PSRs J1326$-$4728A and J1326$-$4728B are shown as blue stars.}
\label{fig:atca}
\end{figure*}

In Fig.~\ref{fig:atca}, we show a $0.1\times0.1$ square degree region centred on the core of \wcen. We overlay unidentified X-ray sources~\citep{hch+18} and the unidentified Fermi source FL8Y J1326.7$-$4729. The core of \wcen\ is shown as the yellow circle with a radius of 155\,arcsec~\citep{har10}. Within the core region we identified several faint continuum sources at 2.1~GHz above a threshold of 3.5$\sigma$ as listed in Table~\ref{tab:atca} and they are shown as purple circles in Fig.~\ref{fig:atca}. The two brightest sources (Nos. 6 and 7) are clearly extended and associated with X-ray
sources, and are likely radio galaxies. Of the remaining sources, only No. 13 is coincident with X-ray emission. In the 5.5~GHz image, only 5 sources are within the core region above a flux density limit of 30~$\mu$Jy. Two are the bright extended sources discussed above, while one corresponds to source No. 2 at 2.1~GHz.
\begin{table}[b!]
\begin{center}
\caption{ATCA sources in the core region of \wcen.}
\label{tab:atca}
\begin{tabular}{lccc}
\hline
\hline
No. & RA & Dec & $S_{2.1}$ \\
& (J2000)	& (J2000) & ($\mu$Jy)    \\
\hline
1 & 13:26:54.37 & $-$47:30:56.3 & 45  \\
2 & 13:26:39.33 & $-$47:30:44.8 & 46  \\
3 & 13:26:46.40 & $-$47:30:29.4 & 55 \\
4 & 13:26:37.26 & $-$47:29:42.1 & 68 \\
5 & 13:26:48.66 & $-$47:29:25.6 & 47  \\
6 & 13:26:41.78 & $-$47:29:19.8 & 135  \\
7 & 13:26:38.15 & $-$47:29:04.8 & 544 \\
8 & 13:26:56.47 & $-$47:28:45.0 & 46 \\
9 & 13:26:56.28 & $-$47:28:30.1 & 58 \\
10 & 13:26:41.74 & $-$47:27:00.4 & 47 \\
11 & 13:26:55.28 & $-$47:26:35.4 & 93 \\
12 & 13:26:42.71 & $-$47:28:46.2 & 40 \\
13 & 13:26:49.57 & $-$47:29:25.0 & 44 \\
14 & 13:26:57.99 & $-$47:30:29.5 & 41 \\
\hline
\end{tabular}
\end{center}
\end{table}

\subsection{Parkes Observations and pulsar details}
\label{sec:pks}
The unidentified Fermi source, FL8Y J1326.7$-$4729, located in the core of \wcen\ was observed on 2018 November 22 and 25 at Parkes using the UWL receiver as part of project P970 (PI: S. Dai). The observation on November 22 used the PDFB4 backend, and data were recorded with 2-bit sampling every 64\,$\mu$s in each of the 0.5\,MHz wide frequency channels (512 channels across the 256\,MHz band centred at 1369\,MHz). The November 25 observation used the Medusa backend, which in conjunction with the UWL, provides a radio frequency coverage from $704$\,MHz to 4032\,MHz~\citep[for details see][]{hmd+19}. Data were recorded with 2-bit sampling every 64\,$\mu$s in each of the 0.125\,MHz wide frequency channels (26624 channels in total). The total integration time was 2700 and 3600\,s, respectively. 

A periodicity search was carried out with the pulsar searching software package {\sc PRESTO}~\citep{ran01}.  The DM range that we searched was $0-500\,{\rm pc\,cm}^{-3}$. In order to account for possible orbital modulation of pulsar periodic signals, we searched for signals drifting by as much as $\pm200/n_{h}$ bins in the Fourier domain by setting $zmax=200$~\citep{rem02}, where $n_{h}$ is the largest harmonic at which a signal is detected (up to 8 harmonics were summed). We identified two pulsars at the same dispersion measure ${\rm DM}=100.3\,{\rm pc\,cm}^{-3}$ with rotation periods of 4.10\,ms and 4.79\,ms, respectively. 

Follow-up observations under the auspices of project P1034 were performed using the coherently de-dispersed search mode where data are recorded with 2-bit sampling every 64\,$\mu$s in each of the 1\,MHz wide frequency channels (3328 channels across the whole band with Medusa). Data were coherently de-dispersed at a DM of 100.3\,${\rm pc\,cm}^{-3}$. Table~\ref{tab:obs} lists the observations and integration times. A periodicity search as described above was carried out for each observation within a DM range of $90-110\,{\rm pc\,cm}^{-3}$. On 2019 November 10, we observed \wcen\ for $\sim10$\,hr. Assuming a bandwidth of 1\,GHz centred at 1.4\,GHz, and with a system equivalent flux density of $\sim36$\,Jy~\citep{hmd+19}, the 10\,hr integration gives us a nominal sensitivity of $\sim20$\,$\mu$Jy for $8\sigma$ detection. We also split the long observation into one-hour blocks and carried out searches for binary systems up to $zmax=200$.

With the apparent spin period of each pulsar determined at each observing epoch, data were folded using the \texttt{DSPSR}~\citep{vb11} software package with a sub-integration length of 30 seconds. We manually excised data affected by narrow-band and impulsive radio-frequency interference for each sub-integration. Each observation was averaged in time to create sub-integrations with a length of a few minutes and pulse time of arrivals (ToAs) were measured for each sub-integration using {\sc psrchive}~\citep{hvm04}. On 2019 October 14, full Stokes information was recorded and a pulsed noise signal injected into the signal path was observed before the observation. Polarisation and flux calibration were carried out for this observation following \citet{dlb+19}.
\begin{table}[b!]
\begin{center}
\caption{Observation summary.}
\label{tab:obs}
\begin{tabular}{lcc}
\hline
\hline
	Start & Epoch & Integration \\
	(UTC) & (MJD) & (second)    \\
\hline
	2018 Nov. 22 & 58444.95 &2691  \\
	2018 Nov. 25 & 58447.83 &3600  \\
	2019 Jun. 9  & 58643.30 &1800  \\
	2019 Jun. 22 & 58656.39 &2960  \\
	2019 Jul. 7  & 58671.49 &2454  \\
	2019 Aug. 5  & 58700.26 &   1768  \\
	2019 Sep. 11 & 58737.10 &   3616  \\
	2019 Oct. 9  & 58765.04 &   3185  \\
	2019 Oct. 10 & 58765.96 &   3405  \\
	2019 Oct. 11 & 58767.05 &   1335  \\
	2019 Oct. 12 & 58767.96 &   8453  \\
	2019 Oct. 13 & 58769.02 &   3592  \\
	2019 Oct. 14 & 58770.01 &   3555  \\
	2019 Oct. 15 & 58771.25 &   1703  \\
	2019 Oct. 29 & 58785.93 &   3295  \\
	2019 Nov. 10 & 58796.79 &   37980 \\
	2019 Nov. 11 & 58798.04 &   3593  \\
\hline
\end{tabular}
\end{center}
\end{table}

\begin{figure*}
\centering
\includegraphics[width=7cm]{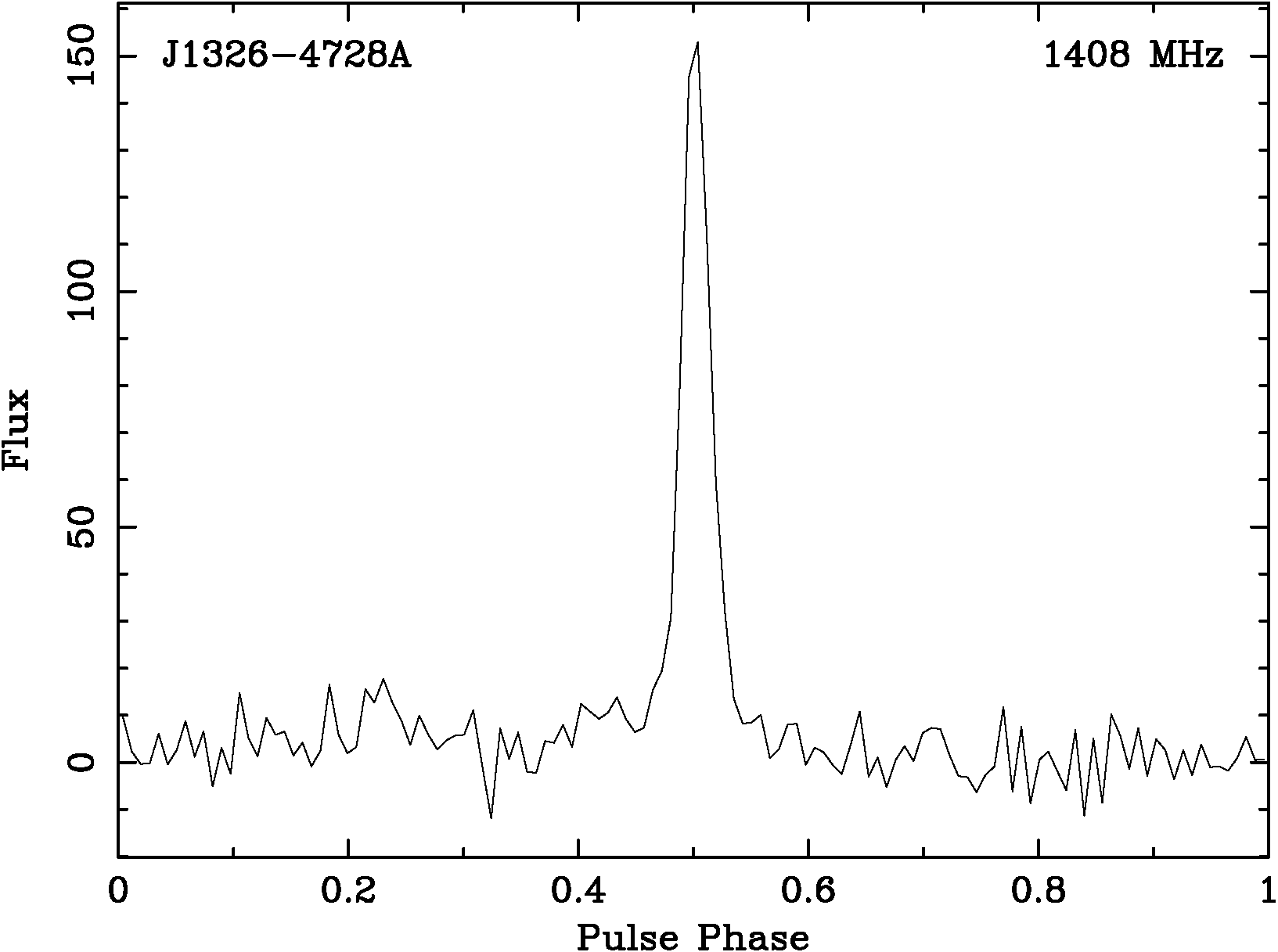}
\includegraphics[width=7cm]{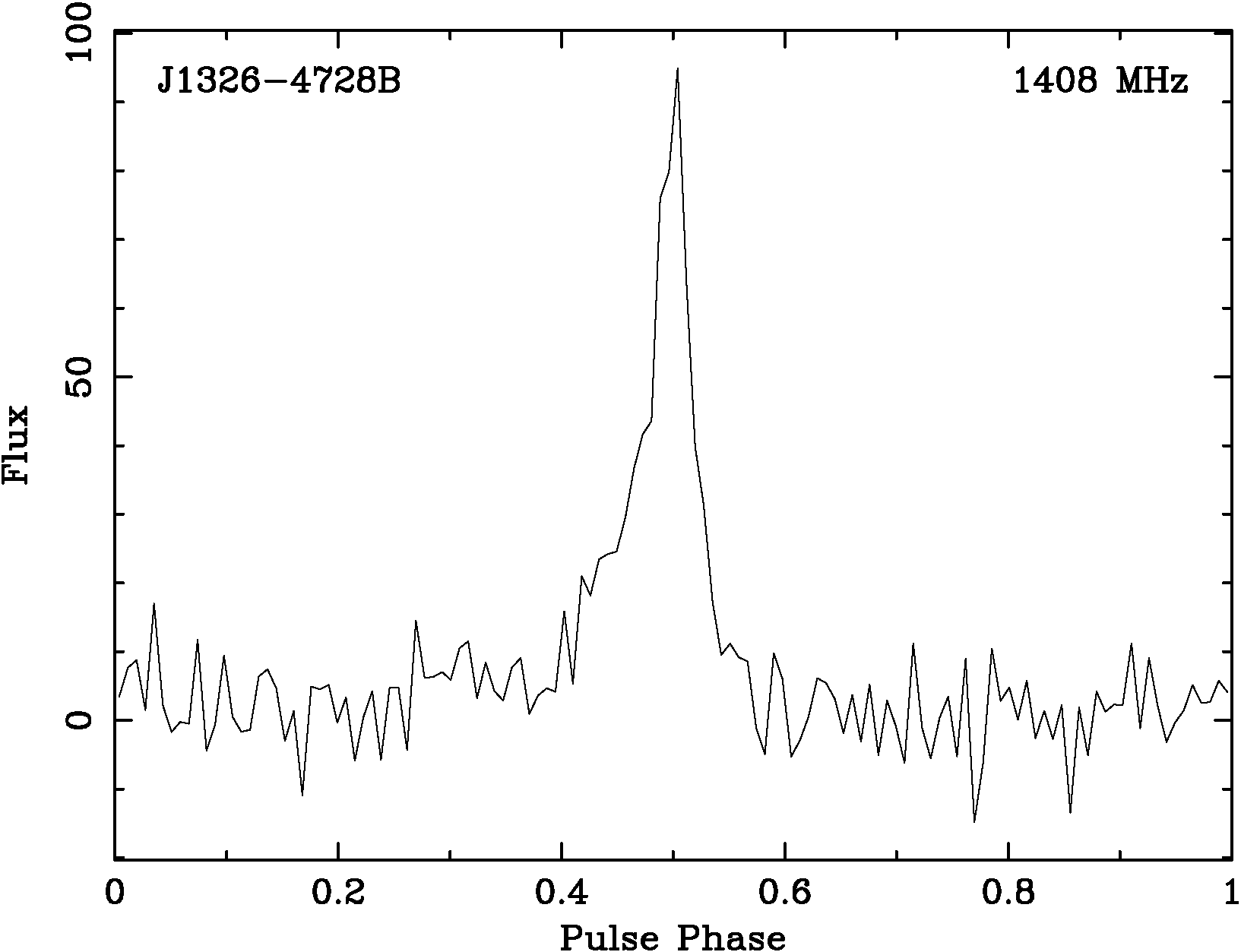}
\includegraphics[width=7cm]{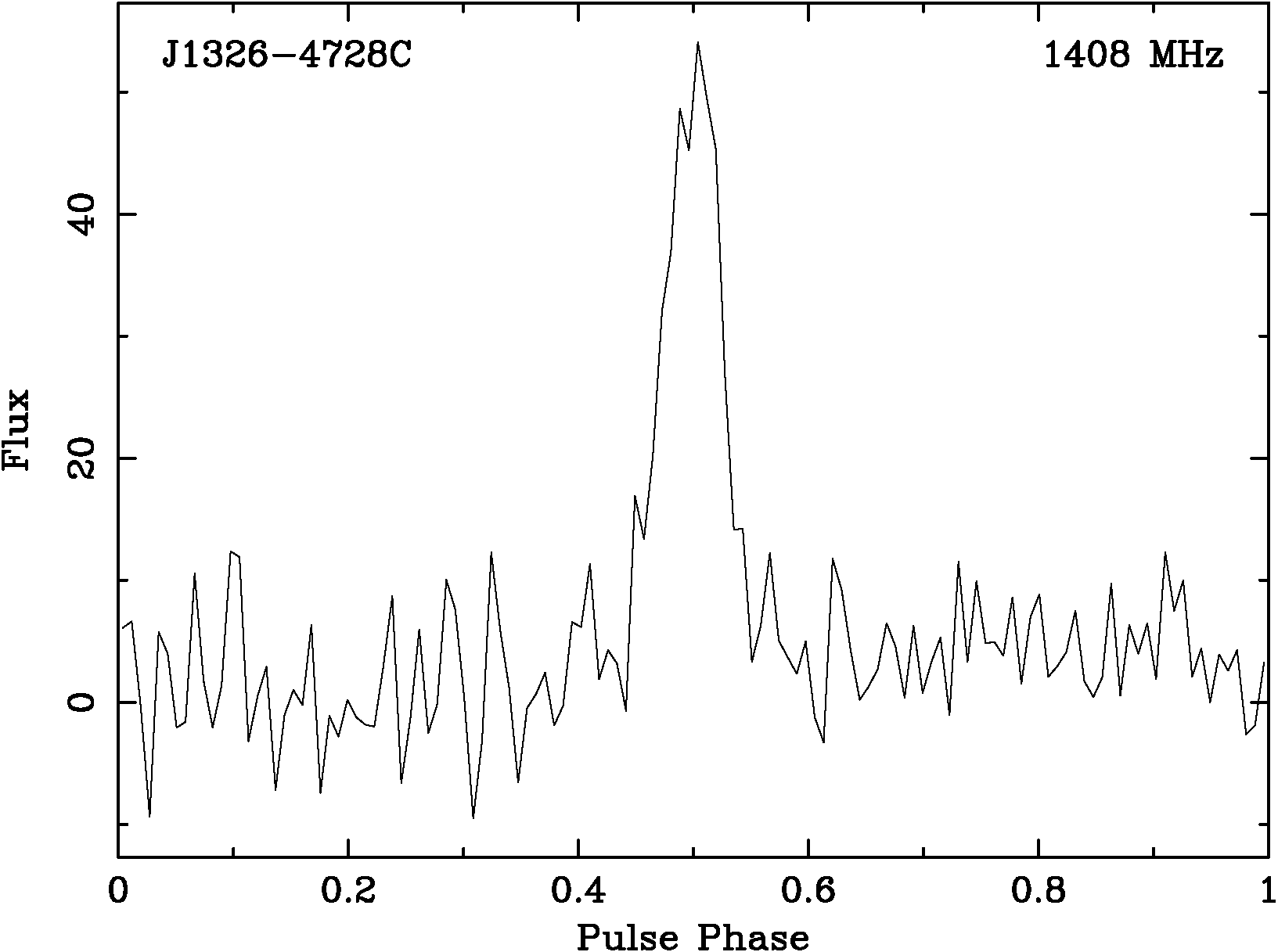}
\includegraphics[width=7cm]{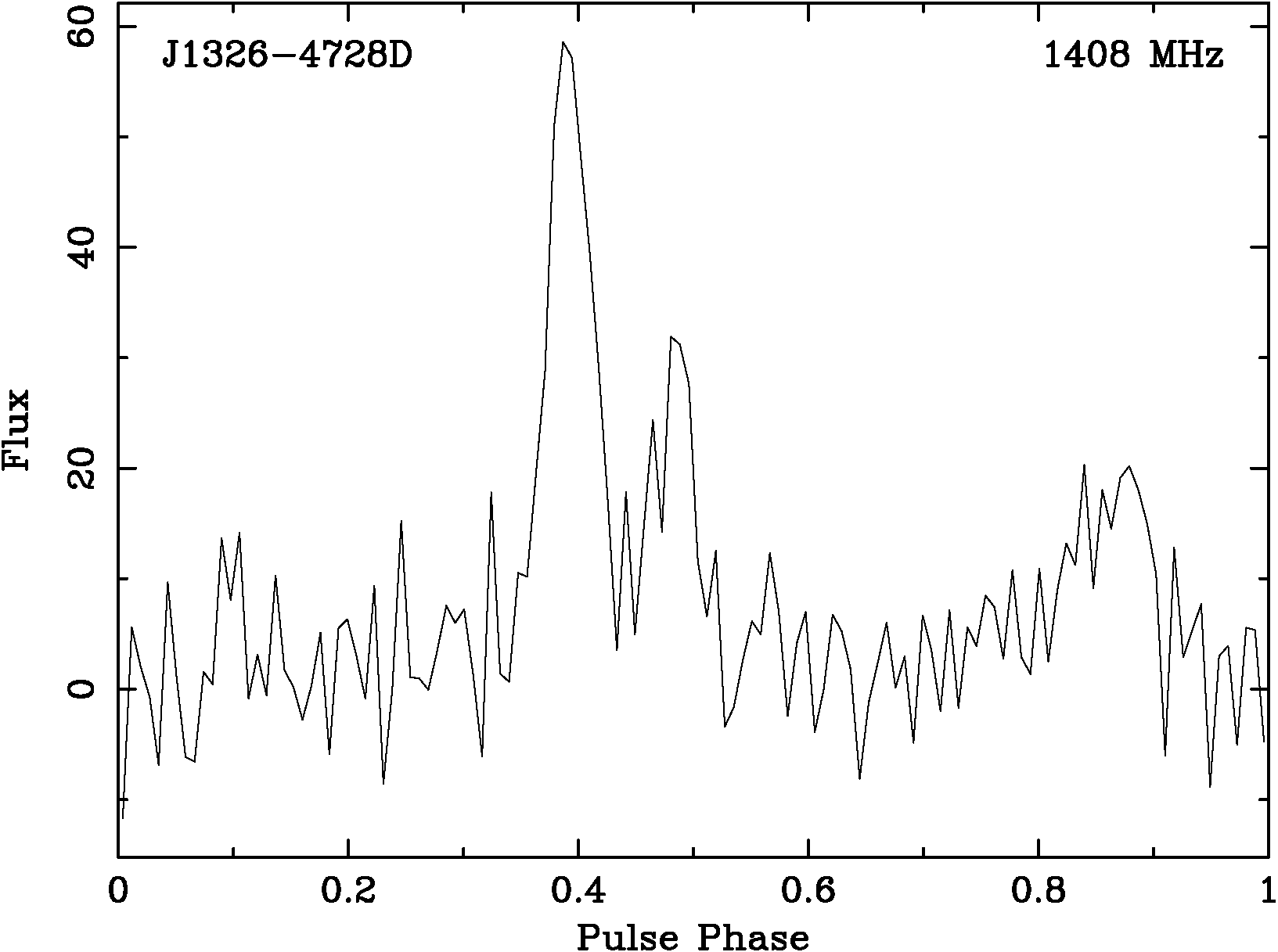}
\includegraphics[width=7cm]{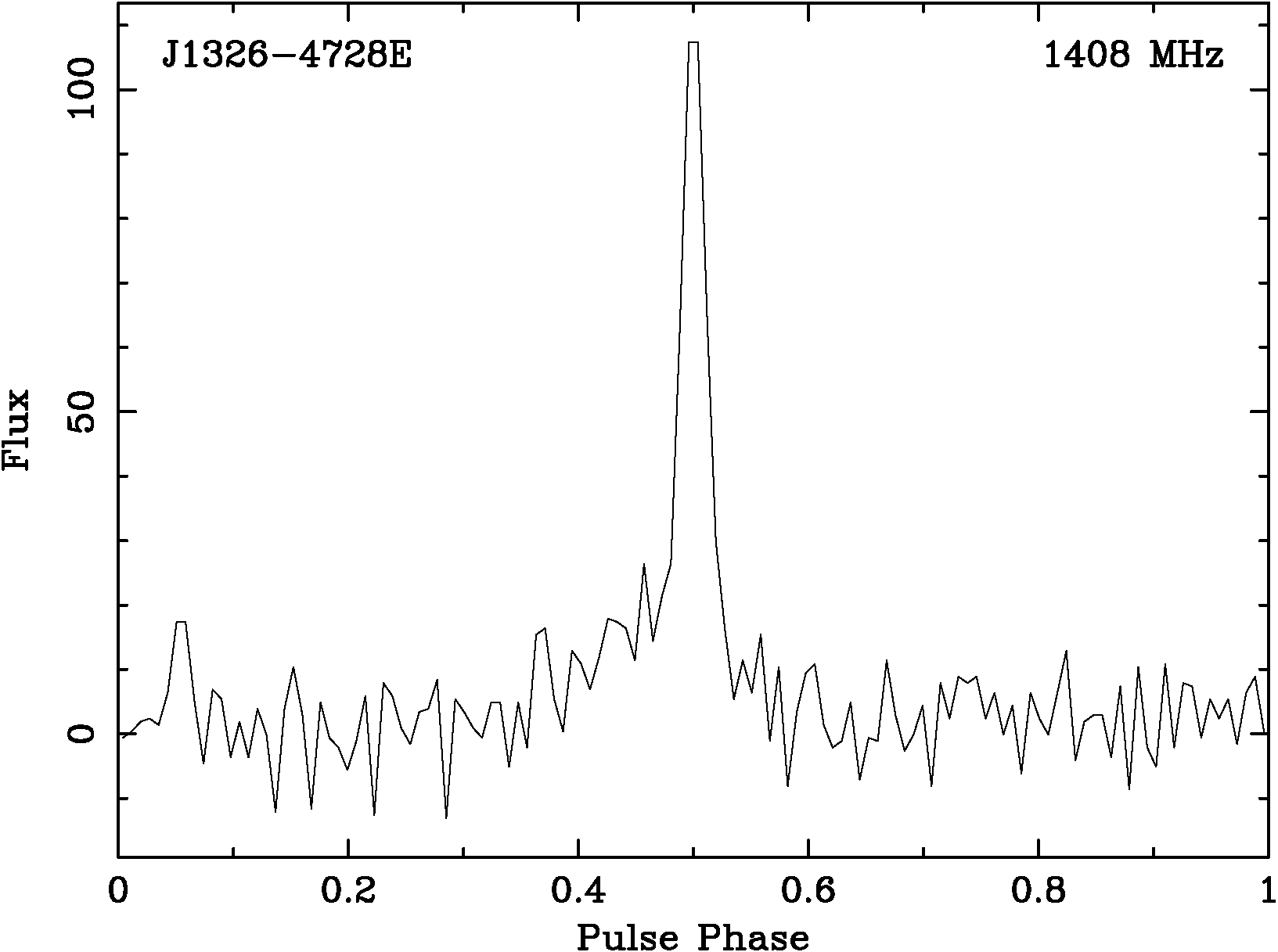}
\caption{Integrated pulse profiles using data taken on 2019 November 10.}
\label{fig:prof}
\end{figure*}

PSR~J1326$-$4728A (4.1\,ms, hereafter pulsar A) shows evidence of strong intensity variability likely caused by interstellar scintillation. The signal-to-noise ratio (S/N) of folded profiles, scaled to one hour integration, varies from 4 to 18. Its integrated profile, shown in Fig.~\ref{fig:prof}, is narrow. The 4.8\,ms pulsar (PSR~J1326$-$4728B, hereafter pulsar B) is in a binary with an orbital period of 0.0896\,days. Clear signs of eclipsing are observed. The pulsar is associated with an ATCA continuum sources (No. 13 in Table~\ref{tab:atca}) and an unidentified X-ray source as shown in Fig.~\ref{fig:atca}. 
During the follow-up campaign, we discovered a 6.8\,ms pulsar (PSR~J1326$-$4728C, hereafter pulsar C), a 4.6\,ms pulsar (PSR~J1326$-$4728D, hereafter pulsar D) and a 4.2\,ms pulsar (PSR~J1326$-$4728E, hereafter pulsar E). These three pulsars are all isolated and significantly fainter than the first two pulsars. They were not detectable in 2018 observations nor short observations in 2019.  

Coherent timing solutions for pulsars A and B were derived from the ToAs via {\sc tempo2}~\citep{hem06} and are listed in Table~\ref{tab:psr}. We note that the values of $\dot{\nu}$ for these pulsars are likely to be contaminated by their acceleration in the gravitational potential of the globular cluster. This also affects the derived values of $B$ and $\dot{E}$ in the table. We have not yet obtained phase-coherent solutions for pulsars C, D and E. DM of each pulsar was measured using sub-band ToAs from 704 to 2112\,MHz of the 10\,hr observation.

Calibrated flux densities and spectral indices were measured with data taken on 2019 October 14 and are given in Table~\ref{tab:psr}. The {\sc psrchive} program {\sc psrflux} was used to measure the flux density. Flux densities were measured from 704 to 2112\,MHz. While pulsars A, C, D and E show steep spectra, the spectrum of pulsar B is flat and it can be detected up to 3\,GHz. We note that spectral indices were 
measured using one observation, and can therefore be affected by the variability of these pulsars and RFI at low frequencies. The spectrum can also be steepened if the pulsar is offset from the pointing centre. For a pulsar with an intrinsic spectral index of -1.5 and offset from the pointing centre by 7\arcmin, the observed spectral index is $\sim-2.3$ within the frequency range from 704 to 2112\,MHz. No linear or circular polarization was detected for any of the pulsars with a limit of 50\%.

\begin{table*}
\begin{center}
\caption{Parameters of the five pulsars.}
\label{tab:psr}
\begin{tabular}{lccccc}
\hline
\hline
          & J1326$-$4728A          &  J1326$-$4728B        & J1326$-$4728C   &  J1326$-$4728D  & J1326$-$4728E\\
\hline 
RAJ (J2000)       & 13:26:39.670(3)         &  13:26:49.563(6)          & 13:26:44\footnote{For pulsars without a timing solution, we quote the Parkes pointing centre as the position.}   & 13:26:44  & 13:26:44     \\
DECJ (J2000)      & $-$47:30:11.64(1)       &  $-$47:29:24.62(2)       & $-$47:29:40 & $-$47:29:40   & $-$47:29:40  \\
$\nu$ (Hz)& 243.380880198(3)        & 208.686833122(5)      & 145.6057622(2)  & 218.39623734(7)   & 237.658566471(2)  \\
$\dot{\nu}$ (Hz/s) & $-1.6(2)\times10^{-15}$ & $-1.2(4)\times10^{-15}$   &    &   & \\
PEPOCH    & 58447.77               &  58768.0      &     58447.77    & 58797.01    &  58796.79 \\
Time span (MJD) &  58444.95$-$58798.08  &  58444.95$-$58798.08  &  &  & \\
DM        & 100.313(3)             &  100.273(3)      &    100.648(4)   & 96.542(3)     &  94.3841(9) \\
$S_{1400}$ ($\mu$Jy)  &  68(7) & 55(5) & 30(4)  &  27(5)  &  19(3) \\
\hline
\multicolumn{6}{c}{Binary parameters~\citep[ELL1 model,][]{hem06}}\\
\hline
$P_{\rm{b}}$ (days) &         &  0.08961121(1)       &         &         \\
$\chi$ (ls) &             &  0.021455(7)         &       &  \\
$T_{\rm{asc}}$ (MJD) &        &  58768.037243(4)     &        &\\
$\eta$ (10$^{-3}$)  &         &  0.1(6)              &       &  \\
$\kappa$ (10$^{-3}$) &       &  -0.4(6)             &       & \\
\hline
\multicolumn{6}{c}{Derived parameters}\\
\hline
$\dot{E}$ (erg s$^{-1}$)  & $1.5\times10^{34}$      &   $1.0\times10^{34}$     &  &  &\\
$B$ (G)  & $3.4\times10^{8}$     &  $3.7\times10^{8}$      &  &  &\\
Spectral index  &   $-1.7(2)$    &  $-0.1(2)$         &  $-2.4(3)$   &   $-2.0(2)$  &  $-3.4(2)$ \\
\hline
\end{tabular}
\end{center}
\end{table*}

\section{Discussion}
\label{sec:discussion}
\wcen\ has been a prime target for pulsar searching. Early searches with the Parkes Multi-beam receiver~\citep{evb01,pdc+05} were pointed at the optical centre of the cluster~\citep{har10}, which is only $\sim1\arcmin$ away from our pointing. More recently, the unidentified Fermi source in \wcen\ was observed by \citet{ckr+15} based on the position in the second Fermi-LAT catalogue, offset from our pointing by $\sim2\arcmin$. Considering the half-power width of the telescope beam at 1.4\,GHz of 14\arcmin, the pointing offset is not the main reason for previous non-detections.
On the other hand, we measured the flux densities of these pulsars to be less than 100\,$\mu$Jy (Table~\ref{tab:psr}).  The nominal (8$\sigma$) sensitivity of previous searches to a 4\,ms pulsar with $\rm{DM}\sim100$\,pc\,cm$^{-3}$ is 0.1 to 0.2\,mJy, higher than the measured flux densities of the new pulsars. Therefore, previous non-detections can be understood by the lack of sensitivity, especially considering the variability of the pulsars and the orbital modulation of pulsar~B. The frequency coverage of UWL, down to 700\,MHz, also greatly improved our sensitivity to pulsars with steep spectra and/or large offsets from the pointing centre.

Pulsars~A and B have similar DMs, but we observe significant differences in DM of 0.3, 3.8 and 5.9\,pc\,cm$^{-3}$ towards pulsars~C, D and E, respectively. Such a DM spread roughly agrees with the linear correlation between DM and DM spreads shown in \citet{fhn+05}. This indicates that the DM variation is dominated by small-scale irregularities in the Galactic electron column density. The distance to \wcen\ has previously been determined to be $\sim5.2$\,kpc through the photometry of RR Lyrae stars~\citep{har10}. 
For this distance, we expect a DM of $\sim91$\,pc\,cm$^{-3}$ based on the YMW16 electron-density model~\citep{ymw17} or $\sim126$\,pc\,cm$^{-3}$ based on the NE2001 model~\citep{cl02}, which bracket our measurements of $94-101$\,pc\,cm$^{-3}$.
The NE2001 model seriously overestimates the amount of scattering in this direction, with an expected scattering time of 8.5~ms at 1~GHz, whereas we see little evidence ($<1$~ms) of scattering even at 700~MHz. 
If we assume a diffractive scintillation bandwidth of $\sim$1\,MHz and a velocity of $\sim$100\,km\,s$^{-1}$ for the MSPs in \wcen\ this would imply a diffractive timescale of $\sim$10\,mins and a refractive timescale of several days. We therefore surmise that refractive scintillation is the main cause of the flux variability seen in the pulsars.

The radio continuum image shows that there are 14 sources within the core of \wcen, some fraction of which are likely to be background radio galaxies. The position of pulsar B coincides with a radio continuum source with a flux density of 44\,$\mu$Jy, in agreement with the Parkes value. No radio continuum source is detected at the timing position of pulsar A. The absence of this pulsar in ATCA images could be due to its strong variability and steep spectrum. The variability of the pulsars means that repeated, long observations of \wcen\ may be required to detect further pulsars~\citep[c.f. the discoveries in 47 Tuc and Terzan 5,][]{phl+16,crf+18}.

Using the orbital parameters determined for pulsar B we can constrain the mass of its binary companion. Assuming a pulsar mass of 1.4\,M$_{\odot}$, the companion mass is estimated to be 0.016\,M$_{\odot}$
for an inclination angle of $i=60^{\circ}$ and the minimum mass 
is 0.0138\,M$_{\odot}$. This suggests that pulsar B is in a `black widow' system similar to those discovered in other GCs~\citep[e.g.,][]{rob13}. While signs of eclipsing have been observed, eclipses seem to be irregular in duration. Similar eclipsing features have been observed in J0024$-$7204V~\citep{clf+00,rft+16}, but in that system the companion is significantly more massive. The observed eclipses and the pulsar's association with an unidentified X-ray source suggest that the pulsar may also be a strong $\gamma$-ray emitter, and considering its distance the X-rays are likely to be produced by intrabinary shocks~\citep[e.g.,][]{grm+14}. Study of this binary system will be the subject of future work.

Given the location of pulsar A and B, their minimum separation from the cluster centre must be 2.8 and 1.1\,pc, respectively. Using a mass for the cluster of $2.8\times10^6$\,M$_{\odot}$~\citep{av10}, we can compute the acceleration towards the cluster centre and hence the maximum acceleration in our line-of-sight. This converts to a maximum $|\dot{\nu}|$ of $1.6\times10^{-14}$ and $8.8\times10^{-14}$\,Hz/s for the two pulsars, considerably higher than the measured values. Therefore either the actual radial acceleration is small, or the pulsars are located on the far side of the cluster centre.

The timing position of pulsars~A and B locate them within the error box of the $\gamma$-ray source FL8Y~J1326.7--4729. As we only have 1~yr of timing data, this is not yet sufficient to detect pulsations from the $\gamma$-ray photons over the 13~yr lifetime of the Fermi mission. However, the detection of five MSPs in \wcen\ strongly suggests that FL8Y~J1326.7--4729 arises from the $\gamma$-ray emission of the MSPs rather than annihilating dark matter. The $\gamma$-ray emission may be dominated by one pulsar as seen in NGC 6624 and M28~\citep{faa+11,whw+13}, and the detection of $\gamma$-ray pulsations will then allow us to put more stringent constraints on dark matter parameters. The black-widow system would seem to be the most likely candidate, given that many of the discovered $\gamma$-ray MSPs are eclipsing binary systems with X-ray emission~\citep{grm+14}. Alternatively, the $\gamma$-ray emission may be the summed emission of the ensemble of MSPs, as is the case for 47~Tuc~\citep{aaa+09}, and so deeper searches for yet more pulsars and measurements of the pulsars' intrinsic $\dot{E}$ will be useful.

\section{Summary}
\label{sec:summary}
We have discovered five MSPs in the direction of an unidentified $\gamma$-ray source coincident with the core of the globular cluster \wcen. All five pulsars have a DM near 100~cm$^{-3}$\,pc, making it almost certain that they reside in the cluster itself. Four of the pulsars are isolated MSPs while the fifth is in an eclipsing binary system, similar to many of the other $\gamma$-ray MSPs. The deep continuum image reveals a small population of sources with flux densities of $\sim$50~$\mu$Jy at 2~GHz, some of which are coincident with X-ray emission. We surmise that further MSPs await discovery. Long term timing of the pulsars in this cluster will aid in our understanding of the Milky Way's most massive cluster and will likely result in the detection of $\gamma$-ray pulsations.

\acknowledgments
We thank J. Green and J. Stevens for scheduling Parkes and ATCA observations. We thank the referee, S. Ransom for improvements to the manuscript. The Australia Telescope Compact Array and the Parkes radio telescope are part of the Australia Telescope National Facility which is funded by the Commonwealth of Australia for operation as a National Facility managed by CSIRO. Work at NRL is supported by NASA.  



\end{document}